\documentclass[aps,prd,reprint,preprintnumbers,superscriptaddress,showpacs,twocolumn]{revtex4-1}
\usepackage{latexsym,graphicx,amssymb,amsmath,mathrsfs}
\usepackage{setspace,bm}
\usepackage[breaklinks, colorlinks=true, pdfstartview=FitV, linkcolor=red, citecolor=blue, urlcolor=blue]{hyperref}

\usepackage{amsmath}
\usepackage[usenames]{color}
\usepackage{latexsym}
\usepackage{epstopdf}

\newcommand{\Slash}[1]{{\ooalign{\hfil/\hfil\crcr$#1$}}}
\newcommand\nc{N_\mathrm{c}}

\newcommand\zc{\mathbb{Z}_{N_\mathrm{c}}}
\newcommand\mur{\mu_\mathrm{R}}
\newcommand\mui{\mu_\mathrm{I}}
\newcommand\trw{T_\mathrm{RW}}
\newcommand\qrw{\theta_\mathrm{RW}}
\usepackage[normalem]{ulem}

\newcommand{\comment}[1]{}

\renewcommand\sout{\bgroup \color{red} \ULdepth=-.5ex \ULset}

\begin{document}
\preprint{YITP-16-38}

\title{Implications of imaginary chemical potential for model building of QCD}

\author{Kouji Kashiwa}
\email[]{kouji.kashiwa@yukawa.kyoto-u.ac.jp}
\affiliation{Yukawa Institute for Theoretical Physics,
Kyoto University, Kyoto 606-8502, Japan}

\begin{abstract}
Properties of QCD at finite imaginary chemical potential are revisited to
 utilize for the model building of QCD in low energy regimes.
For example, the electric holonomy which is closely related to the
 Polyakov-loop drastically affects thermodynamic quantities beside
 the Roberge-Weiss transition line.
To incorporate several properties at finite imaginary chemical
 potential, it is important to introduce the
 holonomy effects to the coupling constant of effective models.
This extension is possible by considering the entanglement vertex.
We show justifications of the entanglement vertex based on the
 derivation of the effective four-fermi interaction in the
 Nambu--Jona-Lasinio model and present its general form with the
 local approximation.
To discuss how to remove model ambiguities in the entanglement
 vertex, we calculate the chiral condensate with different
 $\mathbb{Z}_3$ sectors and the dual quark condensate.
\end{abstract}

\pacs{11.30.Rd, 21.65.Qr, 25.75.Nq}
\maketitle

\section{Introduction}

Understanding non-perturbative properties of Quantum Chromodynamics
(QCD) is one of the important and interesting subjects in nuclear and
elementary particle physics.
The lattice QCD simulation is a promising method to understand such
non-perturbative properties, but the sign problem obscures it at finite
real chemical potential ($\mur$).
Therefore, low energy effective models of QCD are widely used to
investigate the QCD phase structure at finite $\mur$.

In the modeling of QCD in low energy regimes, the chiral condensate
and the Polyakov-loop play an important role to describe the chiral
and confinement-deconfinement transitions.
The famous effective model is the Polyakov-loop extended
Nambu--Jona-Lasinio (PNJL) model~\cite{Fukushima:2003fw}.
In the PNJL model, the quark and gluon contributions are taken
into account through the one-loop level effective potential in usual.
There is the uncertainty how to introduce the gauge boson contribution.
Usually, the chiral part is described by using the Nambu--Jona-Lasinio
(NJL) model or similar quark models in effective models.
However, the gluon part is less clear than the quark part.
There are several effective models for the gluon contribution;
for example,
the logarithmic potential based on  the strong coupling
expansion
\cite{Fukushima:2003fw, Hell:2009by},
the Ginzburg-Landau type potential
\cite{Ratti:2005jh,Roessner:2006xn},
the Meisinger-Miller-Ogilvie model
\cite{Meisinger:2001cq},
the matrix model for the confinement-deconfinement transition
\cite{Dumitru:2010mj,Kashiwa:2012wa},
the effective potential from the Landau-gauge gluon and ghost propagators
\cite{Fukushima:2012qa},
the holonomy potential~\cite{Shuryak:2013tka} and so on.

In addition to individual parts of the quark and gluon contributions,
it is important how to control the strength of the correlation between
the quark and gluon contributions in effective models.
Importance of such correlation can be clearly seen in the
lattice QCD data at finite temperature ($T$) and the imaginary chemical
potential ($\mui$)~\cite{D'Elia:2009qz,Bonati:2010gi}.
At finite $\mui$, there is the first-order transition line which is so
called the Roberg-Weiss (RW) transition and its endpoint which is so
called the RW endpoint~\cite{Roberge:1986mm}.
In Ref.~\cite{D'Elia:2009qz,Bonati:2010gi}, the pion mass ($m_\pi$)
dependence of the RW endpoint has been predicted by the lattice QCD
simulation.
At sufficiently large $m_\pi$, the order
of the RW endpoint is the first-order which is induced by the
first-order $\mathbb{Z}_3$ transition.
The first-order RW endpoint becomes the crossover by the
explicit breaking of the $\mathbb{Z}_3$ symmetry at moderate $m_\pi$.
However, there is the mysterious behavior of the RW endpoint at small
$m_\pi$; the order of the transition turns into the first-order!
This behavior cannot be reproduced by using simple effective model of
QCD, but it can be possible by considering the extension of the
coupling constant.

Recently, it has been proposed in
Ref.~\cite{Kashiwa:2015tna,Kashiwa:2016vrl} that the
confinement-deconfinement transition can be described by using the
analogy of the topological order~\cite{Wen:1989iv} motivated by the
recent progress done in Ref.~\cite{Sato:2007xc} at $T=0$ QCD.
In the conjecture~\cite{Kashiwa:2015tna,Kashiwa:2016vrl}, the nontrivial
free-energy degeneracy at finite $\mui$ plays a crucial role.
The non-trivial degeneracy is related with the drastic change of
the holonomy beside $\theta = (2k-1)k/\nc$ in the deconfinement phase
where $\theta$ is the dimensionless chemical potential
$\theta \equiv \mui/T$, $\nc$ means the number of color and $k$ is any
integer.
Then, we can define the quantum order-parameter which is so called the
quark number holonomy on the manifold in all possible
boundary condition space which is equivalence to the $\theta$
space because the boundary condition and the
imaginary chemical potential have direct relation, see
Ref.~\cite{Roberge:1986mm,Kashiwa:2013rmg} as an example.
From these viewpoints, holonomy effects in the effective model is
crucial not only incorporating the confinement-deconfinement
transition into the model, but also understanding the
confinement-deconfinement transition itself.

In this paper, we revisit the imaginary chemical potential region
to obtain some constraints for the model building of QCD in the low
energy regime.
This paper is organized as follows.
In the next section, we summarize the RW transition and some related
topics on it at finite $\mui$.
Section~\ref{Sec:EV} explains the extension of the effective model based
on Ref.~\cite{Sakai:2010rp}.
Then, we show justifications of the extension based on the derivation of
the effective four-fermi interaction.
In Sec.~\ref{Sec:NR}, we calculate the chiral condensate with different
$\mathbb{Z}_3$ sectors to discuss how to remove ambiguities of model
parameters.
Section~\ref{Sec:Summary} is devoted to summary.

\section{Imaginary chemical potential}

It is well known that there is a special periodicity in the QCD
partition function ($Z$) as a function of $\theta$;
\begin{align}
Z(\theta) &= Z \Bigl(\theta + \frac{2\pi k}{\nc} \Bigr).
\end{align}
This means that several quantities have the $2\pi/\nc$
periodicity.
This periodicity is so called the Roberge-Weiss (RW) periodicity.

In the low and high $T$ region, origins of the RW periodicity is quite
different.
At low $T$, the RW periodicity is realized by only one global minimum on
the complex Polyakov-loop ($\Phi$) plane, but does not at high $T$.
The $\zc$ images should be needed to realize the RW periodicity at high
$T$; for example, the $\mathbb{Z}_3$ images are $e^{i 2\pi/3}$ and
$e^{i4 \pi/3}$ for $\Phi=1$.
This difference of origins seems to be related with the confinement and
the deconfinement nature.
From the different realization of the RW periodicity at low and high
$T$, there should be the first-order transition and its endpoint at
$\theta = (2k-1) \pi / \nc$.
Those are called the RW transition and the RW endpoint, respectively.

When we change $\theta$ with fixing $T$ in the confined phase, the
phase of the Polyakov-loop $\phi$ which is determined from
\begin{align}
\Phi
&= \frac{1}{\nc} \mathrm{tr} {\cal P}
   \Bigl[
   \exp \Bigl( i \oint_0^{\beta} A_4(\tau,{\vec x})~d\tau \Bigr)
   \Bigr]
      = |\Phi| \hspace{0.5mm} e^{i \phi}, 
\end{align}
is continuously rotated as the soliton solution which may be described
by the Jacobi elliptic function where ${\cal P}$ means the path ordering
operator, $g$ is the gauge coupling
constant and $\beta$ means the inverse temperature ($1/T$).
On the other hand, the Polyakov-loop phase is discontinuously
rotated in the deconfined phase.
The discontinuous point appears when we across the RW transition line.
The electric holonomy ($\nu$) is then discontinuously changed beside the RW
transition line and then we obtain
\begin{displaymath}
d \nu = \lim_{\epsilon \to 0}
        \Bigl[ \nu(\qrw-\epsilon) - \nu(\qrw+\epsilon) \Bigr]
\left\{
\begin{array}{l}
       0 ~~~~~~~ T < \trw\\ [5mm]
       \dfrac{2 \pi}{\nc} ~~~~ T > \trw, \\
\end{array}
\right.
\end{displaymath}
where we define $d \nu$ as $\{ d\nu~|~0 \le d \nu \le 2 \pi \}$.
Just on the RW endpoint, $d \nu$ depends on its order:
If the RW endpoint is first-order, $d\nu$ should be $2\pi/\nc$.
Because of this behavior, $\theta$-even quantities such as the entropy
density and pressure should have the cusp, but $\theta$-odd quantities
such as the quark number density have the gap at $\theta=(2k-1)\pi/\nc$
above the RW endpoint.

The RW transition and endpoint have interesting phenomena as discussed above,
but it may have more mysterious behavior which is the triple-point
realization of the RW endpoint at small $m_\pi$ which is predicted by
the lattice QCD simulation~\cite{D'Elia:2009qz,Bonati:2010gi}.
At sufficiently large $m_\pi$, the order
of the RW endpoint is the first-order.
This first-order RW endpoint becomes the crossover by the
explicit breaking of the $\mathbb{Z}_3$ symmetry when we set $m_\pi$ as
moderate values.
However, the order of the RW transition turns into the first-order again
at small $m_\pi$.
This behavior is considered as the consequence of the correlation between
the chiral and the confinement-deconfinement transition nature, but it is
still under debate.
By using simple effective models such as the standard PNJL model, this
behavior cannot be reproduced; for example, see
Ref.~\cite{Sasaki:2011wu}.
To resolve this point, we need extension of the effective model.
Actually, by extending the coupling constant in the PNJL model, we can
reproduce the lattice QCD simulation~\cite{Sakai:2010rp}, but the exact
form of the extended coupling constant is not discussed so far.
In the next section, we revisit this point.

\section{Entanglement vertex}
\label{Sec:EV}

In this study, we use the PNJL model.
The simplest two-flavor and three-color PNJL model Lagrangian density
becomes
\begin{align}
{\cal L}
&= {\bar q} (i\Slash{D}-m_0)q
 + G[ ({\bar q}q)^2
    + ({\bar q} i\gamma_5 {\vec \tau} q)^2]
 - \beta V {\cal U},
\label{Eq:L}
\end{align}
where $D^\mu = \partial^\mu+i \delta^\nu_0 A^{\nu,a} \lambda_a/2$ with
Gell-Man matrices $\lambda_a$,
$m_0$ expresses the current quark mass, $G$ is the coupling
constant and $V$ does the
three-dimensional volume.
The term ${\cal U}$ expresses the gluon contribution.
Model parameters in the NJL part such as $m_0$, $G$ and the cutoff of
the three-dimensional momentum integration ($\Lambda$) is determined by
using the $m_\pi$ and pion decay constant.
The actual values used in this study are
$m_0=5.5$ MeV, $G=5.498$ GeV$^{-2}$, $\Lambda=613.5$ MeV.

In this study, we use the mean-field approximation; see
Ref.~\cite{Sakai:2010rp} for the details of the calculation.
The actual thermodynamic potential can be obtained from the Lagrangian
density (\ref{Eq:L}) as
\begin{align}
\Omega
&= - 4 \int \frac{d^3 p}{(2 \pi)^3}
   \Bigl[ 3 E(p) + T (\ln f^- + \ln f^+) \Bigr]
\nonumber\\
&~~~+ 2 G \sigma^2 + {\cal U},
\end{align}
where
\begin{align}
f^- &= 1 + 3 (\Phi + {\bar \Phi}e^{-\beta E^-(p)}) e^{-\beta E^-(p)}
         + e^{-3 \beta E^-(p)},
\nonumber\\
f^+ &= 1 + 3 ({\bar \Phi} + \Phi e^{-\beta E^+(p)}) e^{-\beta E^+(p)}
         + e^{-3 \beta E^+(p)},
\end{align}
here ${\bar \Phi}$ is the conjugate of $\Phi$ and
$E^\mp(p) = E(p) \mp \mu = \sqrt{p^2 + M^2} \mp \mu$ with
$M = m_0 - 2 G \sigma$ and the chemical potential $\mu=(\mur,\mui)$.
The quantity $\sigma$ becomes the chiral condensate in the present
model.
The gluon contribution ${\cal U}$ is taken into account via the
logarithmic Polyakov-loop effective potential~\cite{Hell:2009by} as
\begin{align}
\frac{{\cal U}}{T^4}
&= - \frac{1}{2} b_2(T) {\bar \Phi} \Phi
\nonumber\\
&+ b_4(T) \ln
     \Bigl[ 1 - 6 {\bar \Phi} \Phi
              + 4 \Bigl( {\bar \Phi}^3 + \Phi^3 \Bigr)
              - 3 ( {\bar \Phi} \Phi )^2
     \Bigr],
\end{align}
where
\begin{align}
b_2(T)
&= a_0
 + a_1 \Bigl( \frac{T_0}{T} \Bigr)
 + a_2 \Bigl( \frac{T_0}{T} \Bigr)^2
 + a_3 \Bigl( \frac{T_0}{T} \Bigr)^3,
\nonumber\\
b_4(T)
&= b_4 \Bigl( \frac{T_0}{T} \Bigr)^3,
\end{align}
with $a_0 = 3.51$, $a_1 = -2.56$, $a_2 = 15.2$, $a_3 = -0.62$ and
$b_4 = -1.68$.
These parameters are determined to reproduce the lattice
QCD data such as the energy and the entropy density in the pure gauge
limit.
The parameter $T_0$ controls the confinement-deconfinement temperature
($T_\mathrm{D}$) in the pure gauge limit and thus we set $T_0 =
270$ MeV because we are interested in the quenched calculation.

In Ref.~\cite{Sakai:2010rp}, the extension of the PNJL model coupling
constant has been proposed.
The extension has been done by the replacement
\begin{align}
G \to G(\Phi)
 = G
   [1 - \alpha_1 {\bar \Phi} \Phi - \alpha_2 ({\bar \Phi}^3 + \Phi^3)],
\label{Eq:EV}
\end{align}
where $\alpha_1$ and $\alpha_2$ are new parameters.
This extended coupling constant is so called the entanglement vertex.
In the extension, the $\mathbb{Z}_3$ symmetry of the vertex is assumed.
This interaction strongly correlates the chiral and the
confinement-deconfinement transition.
The entanglement nature is deeply related with the nonlocal properties
of the interaction which is usually hidden in the local PNJL model.

From the derivation of the NJL and also the PNJL model, the nonlocal
nature is a natural consequence.
The PNJL model can be obtained starting from QCD partition function by
using the color-current expansion and some {\it ansatz} for gluon
properties; details of the nonlocal PNJL model, for example, see
Ref.~\cite{VandenBossche:1997nn,Kondo:2010ts,Kashiwa:2011td} and
references therein.
The nonlocal interaction should feel effects coming from the
gluon two-point function since the interaction forms
\begin{align}
{\cal L}_\mathrm{int} (x)
&= g^2 j_a^\mu (x) \int d^4y~W^{(2) ab}_{\mu\nu} j^\nu_b(y),
\end{align}
where
$j_\mu^a = {\bar q} \frac{\lambda^a}{2} \gamma_\mu q$ is the
color-current, $W^{(2)}$ is the gluon two-point function without quark
loops because we use the color-current expansion around $j = 0$.
Since the two-point function does not have quark loops, $W^{(2)}$ should
be $\zc$ symmetric form.
To obtain the local version of the PNJL model, we should replace the
two-point function as the delta function since we assume $W^{(2)}$ is
very short range.
The gluon follows the adjoint representation of
$SU(\nc)$, and thus the entanglement vertex should obey the adjoint
representation nature.

To introduce the adjoint representation nature to the entanglement vertex,
we refer to the structure of the gluon one-loop effective potential and
the Bose distribution function with the electric holonomy.
Those manifest the $\mathbb{Z}_3$ symmetric form and other adjoint
representation natures.
So, more general form of the entanglement vertex for the color singlet
sector can be written as
\begin{align}
G(\Phi) &= G_\mathrm{N}
    \Bigl[ 1
         + \Bigl(
           \sum_{n=1}^8 \alpha'_n C_n e^{-n \beta E}
           \Bigr)^{\gamma}
    \Bigr],
\label{Eq:EV_a}
\end{align}
where $G_\mathrm{N}$, $\alpha'_n$ and $\gamma$ are parameters,
$E = |{\vec p}|$ is the typical energy scale in the momentum space
and
\begin{align}
C_1 &= C_7 = 1 -9 {\bar \Phi} \Phi,
\nonumber\\
C_2 &= C_6 = 1 - 27 {\bar \Phi} \Phi + 27 ({\bar\Phi}^3+\Phi^3),
\nonumber\\
C_3 &= C_5 = - 2 + 27 {\bar \Phi} \Phi - 81 ({\bar \Phi}\Phi)^2,
\nonumber\\
C_4 &= 2 \Bigl[ - 1 + 9 {\bar \Phi}\Phi - 27 ({\bar \Phi^3 + \Phi^3})
                + 81 ({\bar \Phi}\Phi)^2 \Bigr],
\nonumber\\
C_8 &= 1.
\end{align}
For details of functions $C_n$, see Ref.~\cite{Sasaki:2012bi}.
Even if we assume the effective gluon mass such as the Gribov-Stingl
form at finite $T$ as
\begin{align}
D^{(T,L)}_T
&\propto \frac{d_{t,l}(p^2+d_{t,l}^{-1})}{(p^2+r^2_{t,l})^2},
\end{align}
we then obtain same expression of $C_n$~\cite{Fukushima:2012qa},
where $D^{T}_T$ ($D^{L}_T$) are the tree-dimensional transverse
(longitudinal) gluon propagator.
Thus, we can expect that above expression is general form for the
entanglement vertex.

When we assume that the $W^{(2)}$ is short range, higher-order terms of
$e^{-\beta E}$ in Eq.~(\ref{Eq:EV_a}) should be suppressed
and then only $C_1$ is relevant.
From here, we set $\gamma=1$ to make our discussion simple and it is not
an essential point in the level of present discussions.
Thus, we can effectively rewrite the coupling constant (\ref{Eq:EV_a}) as
the simple form
\begin{align}
G(\Phi) &\sim G_\mathrm{N} \Bigl( 1 + \alpha {\bar \Phi} \Phi \Bigr),
\end{align}
where $G_\mathrm{N}$ is corresponding to $G$ in the original PNJL
model in this case because ${\bar \Phi}$ and $\Phi$ do not affect the
physics at $T=0$.
This form is similar to Eq.~(\ref{Eq:EV}), but it is more general.
If we set $\alpha = - \alpha_1$ and $\alpha_2 =0$ with $\gamma=1$, this
vertex is perfectly matched with the vertex (\ref{Eq:EV}).
From present representation, terms proportional to
$({\bar \Phi}^3 + \Phi^3)$
becomes the higher-order contribution and thus it can be safely
neglected in the level discussed here.
In the effective model construction, important operation is the
parameter fixing.
Details of parameter fixing are discussed in the next section.

Recently, it has been numerically shown that the $\mathbb{Z}_3$ factor
can affect the Landau-gauge gluon propagator in the deconfined phase by
using the pure gauge calculation in Ref.~\cite{Silva:2016onh}.
In the gluon propagator, the $\mathbb{Z}_3$ factors can appear via
${\cal A}_4^{ab} = {\cal A}_4^a - {\cal A}_4^b$ when we consider the
background gauge field where ${\cal A}_4$ is the temporal component of the
field and $a,b$ run $1,\cdots,\nc$ because the gluon obeys the adjoint
representation.
For $a=b$ diagonal sectors, the $\zc$ factor cannot appear
because of the exact cancellation of the $\mathbb{Z}_3$ factors.
On the other hand, the non-diagonal $a \neq b$ sectors may feel the
effect via the effective gluon mass.
Also, in the confined and deconfined phases, the elementary degree of
freedom is drastically different; it is glueball in the confined phase,
but it is gluon in the deconfined phase.
Both reasons may lead to the lattice QCD predicted phenomena, but we
need more careful investigation for the lattice QCD data to conclude it.
In the short range limit of $W^{(2)}$, the color structure is extremely
simplified and thus above effect cannot appears in the local PNJL model.
To incorporate the effects, we should consider the nonlocal PNJL model,
but it seems a difficult task.
Even in the modern nonlocal PNJL
model~\cite{Hell:2008cc,Hell:2009by,Radzhabov:2010dd,Pagura:2011rt,Kashiwa:2011td,Carlomagno:2013ona},
we implicitly use
the simplification of the color structure of the four-fermi interaction.
At present, we do not know how important this effects for the gauge
invariant observable, and thus
we assume that $W^{(2)}$ is short range to neglect the effect as a first
step.
This problem will be discussed in the nonlocal PNJL model in elsewhere.

\section{Numerical results}
\label{Sec:NR}

In this section, we discuss how to remove ambiguities of the
parameters in the entanglement vertex by
using lattice QCD simulations with the quenched approximation.
The entanglement vertex correlates the chiral and
the confinement-deconfinement transition and thus the chiral condensate
with different $\mathbb{Z}_3$ sectors seems to be a promising quantity
to remove ambiguities of model parameters.
The existence of non-trivial $\mathbb{Z}_3$ sectors which is
characterized by $\phi = \frac{2\pi}{3}$ and $\frac{4\pi}{3}$ is not
trivial in the system with dynamical quarks.
Therefore, we here consider the quenched approximation since all
possible $\mathbb{Z}_3$ sectors are well defined.
The calculation of the chiral condensate with the different
$\mathbb{Z}_3$ sector was done in
Ref.~\cite{Chandrasekharan:1995gt,Chandrasekharan:1995nf} and thus we
follow their calculation by using present extended model.
In the calculation~\cite{Chandrasekharan:1995nf} , they introduce the
$T$-dependent coupling constant by hand, but this effect is automatically
introduced in the present entanglement PNJL model.

\begin{figure}[t]
\begin{center}
 \includegraphics[width=0.4\textwidth]{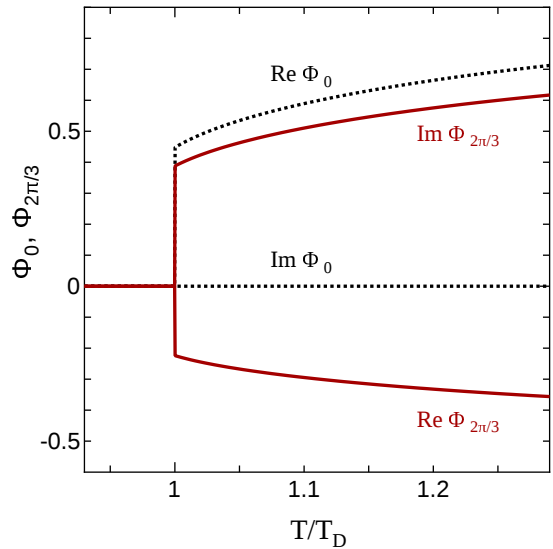}
\end{center}
\caption{
The $T$-dependence of $\Phi$ with different $\mathbb{Z}_3$ sectors in
 the pure gauge limit.
The dotted and solid lines represent the result with $\phi=0$ and
 $\phi=\frac{2\pi}{3}$, respectively.
}
\label{Fig:Pol}
\end{figure}

In this study, we first calculate $\Phi$ in the pure
gauge limit to consider the quenched calculation.
The Polyakov-loop with non-trivial $\mathbb{Z}_3$ sectors,
$\Phi_{\frac{2\pi}{3}}$ and $\Phi_{\frac{4\pi}{3}}$,
are obtained by the $\mathbb{Z}_3$ transformation from
the Polyakov-loop with the trivial $\mathbb{Z}_3$ sector, $\Phi_0$.
Figure~\ref{Fig:Pol} shows $\Phi$ with $\phi=0$ and
$\phi=\frac{2\pi}{3}$ in the pure gauge limit as a function of $T$.
After this calculation, we substitute those $\Phi$
to the PNJL model and then we minimize the thermodynamic
potential as a function of $\sigma$.
This procedure is corresponding to the quenched calculation of the PNJL
model.

\begin{figure}[t]
\begin{center}
 \includegraphics[width=0.4\textwidth]{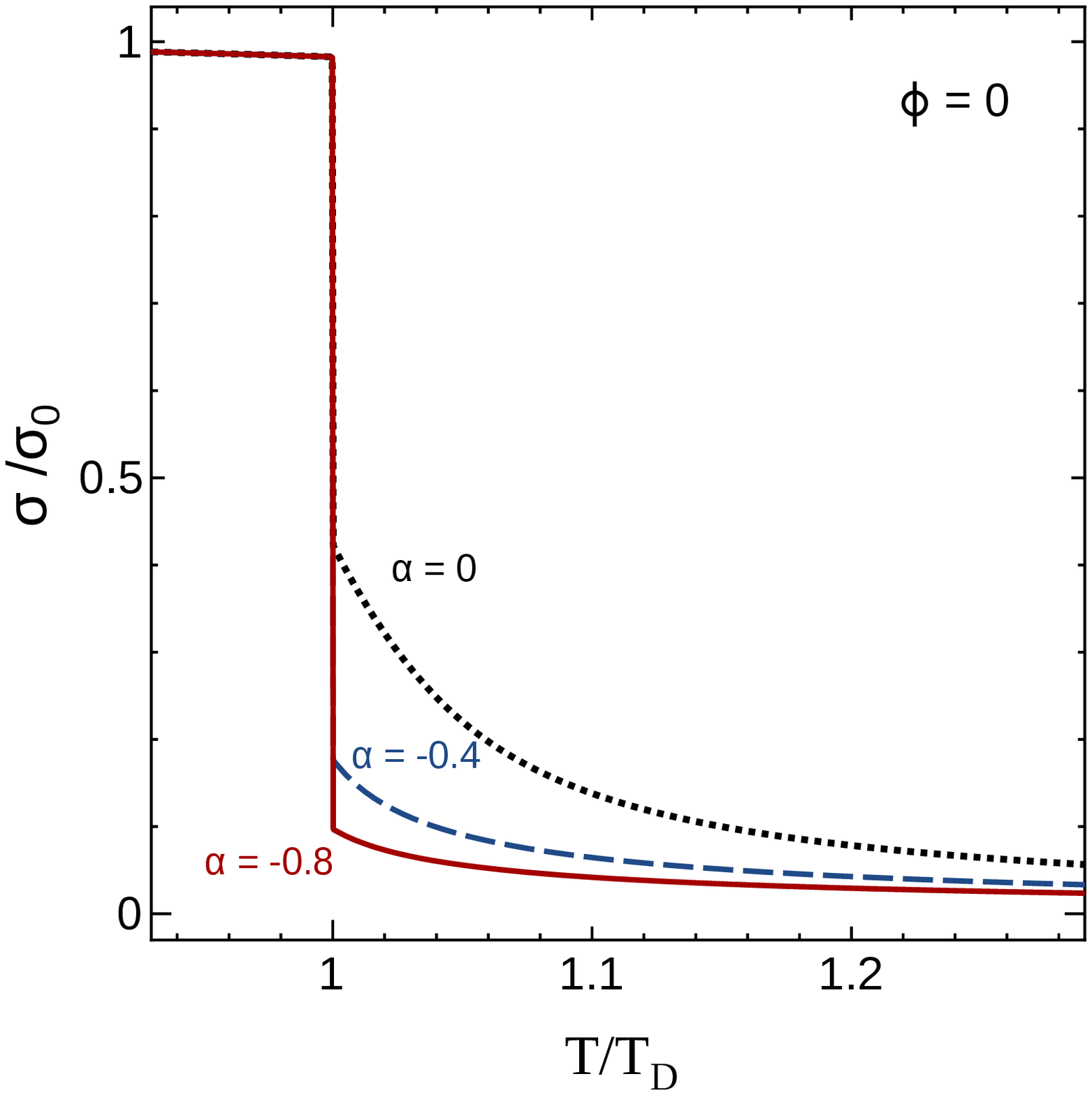}\\
 \vspace{5mm}
 \includegraphics[width=0.4\textwidth]{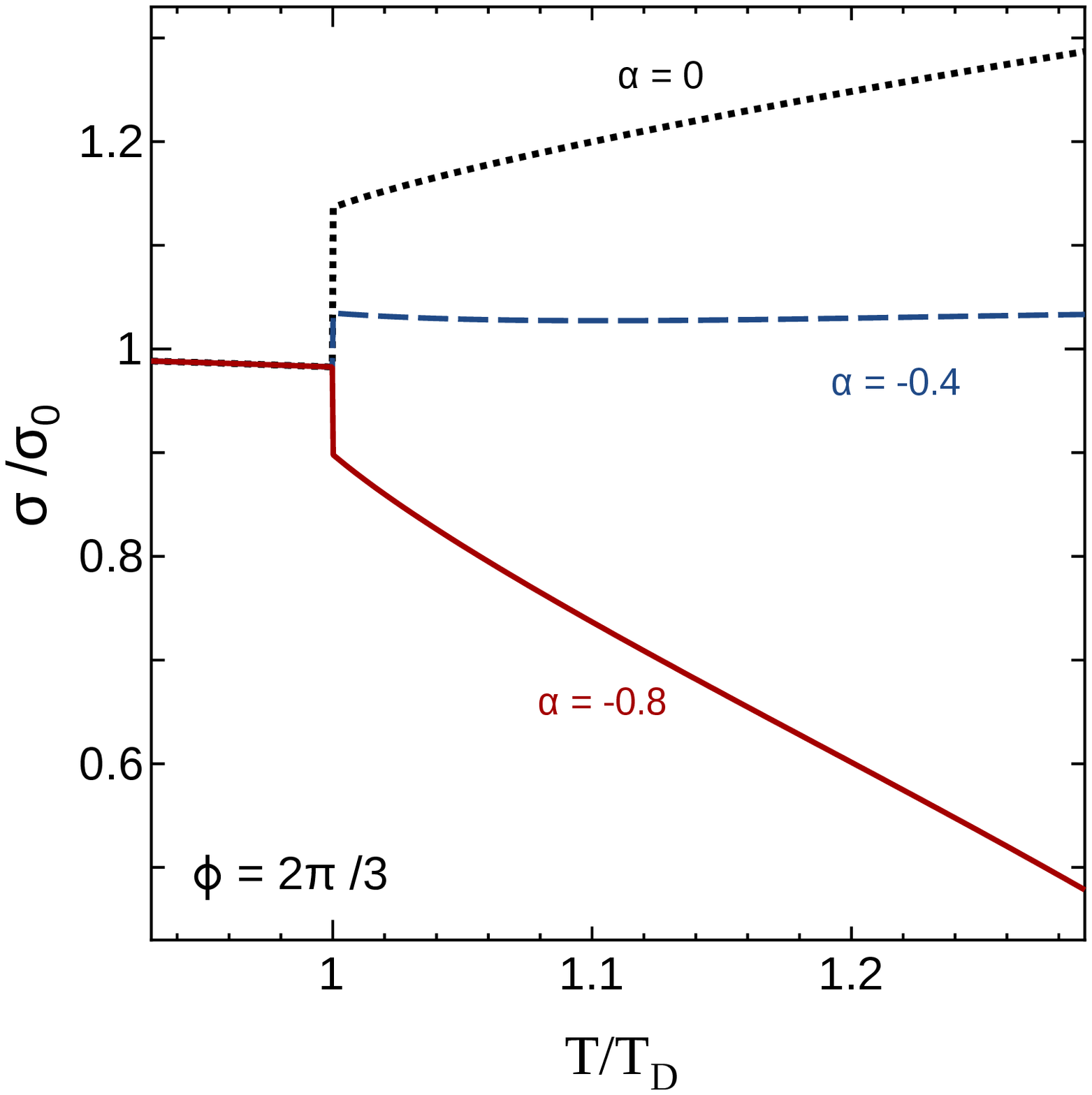}
\end{center}
\caption{
The $T$-dependence of the chiral condensate with different
 $\mathbb{Z}_3$ sectors normalized by own value at $T=\mu=0$.
The top and bottom panels show $\sigma$ with $\phi = 0$ and
 that $\phi =2\pi/3$, respectively.
The dotted, dashed and solid lines represent the result with $\alpha=0$,
 $-0.4$ and $-0.8$, respectively.}
\label{Fig:CC_Z3}
\end{figure}

Figure.~\ref{Fig:CC_Z3} shows the chiral condensate with different
$\mathbb{Z}_3$ sectors at $\mu=0$ as a function of $T$.
The top and bottom panels of Fig.~\ref{Fig:CC_Z3} show $\sigma/\sigma_0$
with the trivial $\mathbb{Z}_3$ sector $\phi=0$ and that with
$\phi = 2 \pi /3$ where $\sigma_0$ is $\sigma$ at $T=\mu=0$,
respectively.
It is natural to assume that the entanglement vertex is decreased with
increasing $T$ and thus the sign of $\alpha$ is negative
because of properties of the running coupling in QCD.
When $\alpha$ is small in the bottom panel, the chiral condensate
increases above $T_\mathrm{D}$, but it decreases in the large-$\alpha$
case.
There is the drastic change in the $T$-dependence of $\sigma$ and thus
we can determine the parameters in the entanglement
vertex by using precise quenched lattice QCD data.
At least in Ref.~\cite{Chandrasekharan:1995gt,Chandrasekharan:1995nf},
the chiral condensate with nontrivial $\mathbb{Z}_3$ sectors are
decreased and thus $\alpha$ should be considerably large.
To conclude the detailed value of $\alpha$, we need more precise
quenched lattice QCD data.

\begin{figure}[t]
\begin{center}
 \includegraphics[width=0.4\textwidth]{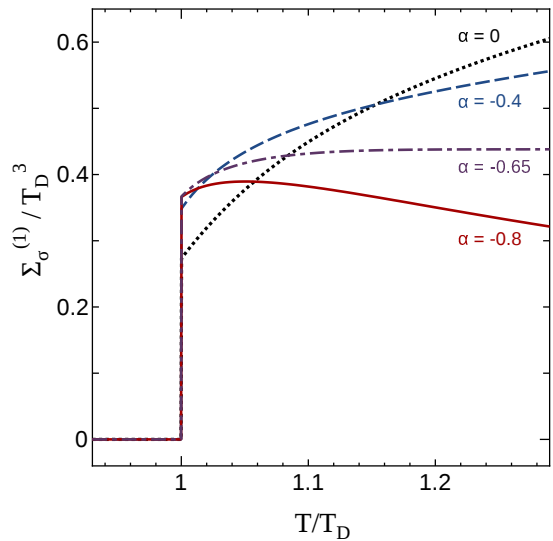}
\end{center}
\caption{
The $T$-dependence of $\Sigma_\sigma^{(1)}$ at $\mur=0$.
The dotted, dashed, dot-dashed and solid lines represent the result with
 $\alpha=0$, $-0.4$, $-0.65$ and $-0.8$, respectively.
}
\label{Fig:DQC}
\end{figure}

Other promising quantity to remove the model ambiguities is the dual
quark condensate~\cite{Bilgici:2008qy,Bilgici:2009tx,Bilgici:2009phd}.
The dual quark condensate is the order-parameter
of the $\mathbb{Z}_{N_\mathrm{c}}$ symmetry breaking as same
as the Polyakov-loop.
This quantity has been calculated by using the lattice QCD
simulation~\cite{Bilgici:2008qy,Bilgici:2009tx,Bilgici:2009phd},
the Schwinger-Dyson equations~\cite{Fischer:2009wc},
the effective model of QCD~\cite{Kashiwa:2009ki},
the functional renormalization group equations~\cite{Braun:2009gm}
and so on.
The dual quark condensate is defined as
\begin{align}
\Sigma^{(n)}_\sigma
&= - \int \frac{d \varphi}{2 \pi}
     e^{- i n \varphi} \sigma_\varphi,
\label{Eq:DQC}
\end{align}
where $\varphi$ specifies the boundary condition of the temporal
direction for quarks, $n$ means the winding number for the temporal
direction and
$\sigma$ is the $\varphi$-dependent chiral condensate.
Usually, we take $n=1$.
The boundary condition and the imaginary chemical potential have
direct relation, $\theta = \varphi - \pi$.
In the system with dynamical quarks, there are some problems in the dual
quark condensate~\cite{Benic:2013zaa,Marquez:2015bca,Zhang:2015baa}, but
we can expect that the quenched calculations where the dual quark
condensate is well defined can provide the important
information for the effective model construction.
At present, quenched lattice QCD data for the dual quark condensate
are limited and thus we do not quantitatively compare our results with
the lattice data.
Figure~\ref{Fig:DQC} shows the $T$-dependence of $\Sigma_\sigma^{(1)}$
at $\mu=0$ with different values of $\alpha$.
We can find the visible $\alpha$-dependence in $\Sigma_\sigma^{(1)}$.
At least in quenched lattice QCD data we can
access~\cite{Bilgici:2008qy},
$\Sigma_\sigma^{(1)}$ increases with increasing $T$ and thus extremely
strong $\alpha$ may be excluded, but we need more quenched lattice QCD
data to conclude it.
The critical value ($\alpha_c$) which leads to the flat behavior of
$\Sigma_\sigma^{(1)}$ above $T_\mathrm{D}$ exists in the range
$-0.7 < \alpha_c < -0.6$.

\begin{figure}[t]
\begin{center}
 \includegraphics[width=0.4\textwidth]{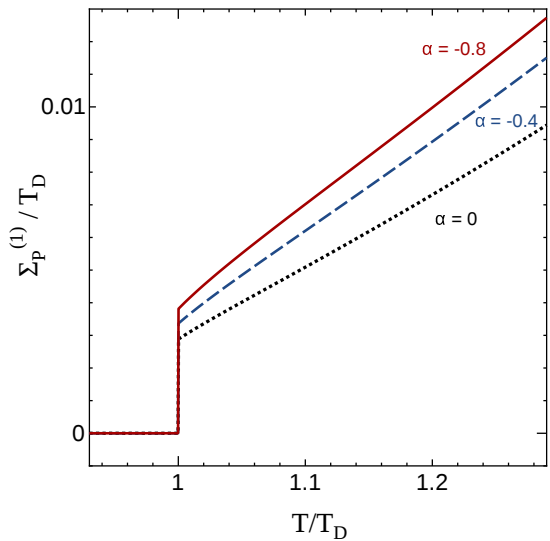}
\end{center}
\caption{
The $T$-dependence of $\Sigma_P^{(1)}$.
The dotted, dashed and solid lines represent the result with $\alpha=0$,
 $-0.4$ and $-0.8$, respectively.
}
\label{Fig:DP}
\end{figure}

Finally, we propose a new quantity which can describe the
confinement-deconfinement transition based on the dual quark condensate.
The quantity is defined as
\begin{align}
\Sigma^{(n)}_{\cal O}
&= \int \frac{d \varphi}{2 \pi}
   e^{- i n \varphi} {\cal O}(\varphi),
\label{Eq:DQ}
\end{align}
where ${\cal O}$ is the $\theta$-even quantity.
If we consider that the dual quark condensate measures how
information of $\mathbb{Z}_3$ images are missed in the calculation of
Eq.~(\ref{Eq:DQC}) when the gauge configuration is fixed to the trivial
$\mathbb{Z}_3$ sector, there is no need to use $\sigma$ as the
integrand of Eq.~(\ref{Eq:DQC}).
It is related with the study based on the nontrivial free energy
degeneracy in Ref.~\cite{Kashiwa:2015tna,Kashiwa:2016vrl}.
Figure~\ref{Fig:DP} shows $T$-dependence of the quantity~(\ref{Eq:DQ})
with ${\cal O} = P$ and $n=1$ where $P$ is the pressure.
We can see that this quantity also describes the
confinement-deconfinement transition.
This behavior is, of course, trivial from the viewpoint of the RW
periodicity.
It is interesting to calculate it in the system with dynamical quarks
and some theories where we cannot easily calculate the chiral condensate.

\section{Summary}
\label{Sec:Summary}

In this paper, we have discussed the extension of the effective model
such as the Polyakov-loop extended Nambu--Jona-Lasinio model.
Firstly, implications of QCD properties in the imaginary chemical
potential region to the effective model construction has been discussed.
Then the importance of the entanglement vertex~\cite{Sakai:2010rp}
has been explained from the derivation of the effective four-fermi
interaction from QCD partition function with some approximations and
{\it ansatz}.

Secondly, we have shown that the general form of the entanglement vertex
which should be needed to reproduce the lattice QCD prediction at finite
$\mui$~\cite{D'Elia:2009qz,Bonati:2010gi}.
The general form is refer to the structure of the gluon one-loop
effective potential and the Bose distribution function and then it
naturally leads to the $\mathbb{Z}_3$ symmetric form.

Finally, the chiral condensate with different $\mathbb{Z}_3$ sectors
and the dual quark condensate with $n=1$
have been calculated by using the PNJL model with the entanglement
vertex.
The entanglement vertex has one parameter $\alpha$ when we assume
$\gamma=1$ at least in the next leading-order of $e^{-\beta E}$.
When $\alpha$ is mall, the chiral condensate with the non-trivial
$\mathbb{Z}_3$ sector increases even above $T_\mathrm{D}$, but it
decreases in the large-$\alpha$ case.
This result suggests that we can well remove the ambiguity of the
parameters by using the $T$-dependence of the chiral condensate
with the different $\mathbb{Z}_3$ sectors.
Also, we can find visible $\alpha$-dependence in the dual quark
condensate and thus this quantity should be the promising quantity to
remove the ambiguity of the parameters.
From qualitative behavior of quenched lattice QCD
data~\cite{Chandrasekharan:1995gt,Chandrasekharan:1995nf,Bilgici:2008qy},
we may exclude extremely strong and weak $\alpha$ values, but we need more
quenched lattice QCD data to conclude it.
In addition to above results, we propose new quantity which can
describe the confinement-deconfinement transition based on the dual
quark condensate.

\begin{acknowledgments}
K.K. thanks A. Ohnishi and H. Kouno for useful comments.
K.K. is supported by Grants-in-Aid for Japan Society for the Promotion
 of Science (JSPS) fellows No.26-1717.
\end{acknowledgments}

\bibliography{ref.bib}

\end{document}